\documentclass[aps,pra,twocolumn,a4paper,nofootinbib,preprintnumbers] {revtex4-1}
\usepackage{graphicx, color}
\usepackage[dvipsnames]{xcolor}
\usepackage[colorlinks=true,urlcolor=blue,citecolor=blue,linkcolor=blue]{hyperref}
\usepackage{amsmath}
\usepackage{amssymb}
\usepackage{amsbsy}

\newcommand{\D}{{\rm{d}}}
\newcommand{\I}{{\rm{I}}}

\begin{document}
\preprint{PHYSICAL REVIEW A \textbf{97}, 063852 (2018)}
\title{Theory of atmospheric quantum channels based on the law of total probability}

 \author{D. Vasylyev}
 \affiliation{Institut f\"ur Physik, Universit\"at Rostock,
 Albert-Einstein-Stra\ss e 23, 18059 Rostock, Germany}

 \author{W. Vogel}
 \affiliation{Institut f\"ur Physik, Universit\"at Rostock,
  Albert-Einstein-Stra\ss e 23, 18059 Rostock, Germany}

 \author{ A. A. Semenov}
\affiliation{Institut f\"ur Physik, Universit\"at Rostock,
 Albert-Einstein-Stra\ss e 23, 18059 Rostock, Germany}

\begin{abstract}
	The atmospheric turbulence is the main factor that influences quantum properties of propagating optical signals  and may sufficiently degrade the performance of quantum communication protocols.
	The probability distribution of transmittance (PDT) for free-space channels is the main characteristics of the atmospheric links.
	Applying the law of total probability, we derive the PDT by separating the contributions from turbulence-induced beam wandering and beam-spot distortions.
	As a result, the obtained PDT varies from log-negative Weibull  to truncated log-normal distributions depending on the channel characteristics.
	Moreover, we show that the method allows one to consistently describe beam tracking, a procedure which is  typically used in practical long-distance free-space quantum communication.
	We analyze the security of decoy-state quantum key exchange through the turbulent atmosphere and show that beam tracking does not always improves quantum communication.
\end{abstract}

\maketitle


\section{Introduction}

	In conventional optical communication the optical signal is transmitted through optical fibers or free space links over hundreds of kilometers.
	In the latter case, the channels are mobile, do not require  access to the optical-fiber infrastructure, and have the potential to establish global quantum communication via satellites.
	The recent  studies of short-distance intracity \cite{Resch, Martinez, Peuntinger, Krenn, Croal, Vasylyev2017}, long-distance ground-based~\cite{Manderbach, Fedrizzi2009, Yin, Ma2012,  Capraro, Herbst, Krenn2016}, and satellite-mediated~\cite{Rarity, Scott, Wang, Bourgoin, Vallone2015, Vallone2016, Yin2017, Guenthner, Liao, GangRen, Takenaka, Liao2018}  quantum links   have shown that quantum protocols are feasible even if the quantum signal undergoes large losses.
	The proper tracking, postselection, preselection, and adaptive strategies~\cite{Elser, Erven, Usenko, Semenov2012, Bourgoin2015, Gruneisen, Bohmann2016} could even improve the performance of the free-space channel.
	Hence, the free-space quantum links are prospective channels for performing quantum key distribution, quantum teleportation, quantum sensing, etc.

	The main obstacles for the efficient performance of quantum protocols with free-space quantum channels are atmospheric turbulence, random scattering, and absorption losses.
	The absorption and scattering effects contribute merely to energy losses and the degradation of the signal intensity.
	An optical beam  that carries a quantum signal also undergoes  amplitude  and phase fluctuations due to the random distribution  of the atmospheric refractive  index.
	The random variation of the refractive index is turbulent in its nature and originates from disordered   mixing of air layers with different temperature, pressure, and humidity characteristics~\cite{Tatarskii, Andrews}.
	Turbulent air motion represents a set of air blobs and vortices and spans a wide range of scales ranging from extremely large to very small.
	Since the optical signal interacts along the propagation path with almost the  whole set of scales, the precise description of light propagation in turbulence is almost impossible and the free-space channel should be described by statistical means.
	In this context the probability distribution of atmospheric transmittance (PDT) through the atmospheric channels, which characterizes the optical channel~\cite{Vasylyev2012, Vasylyev2016}, plays a crucial role.

	In  typical communication scenarios via quantum atmospheric channels the sender generates a signal and sends it through the atmospheric link.
	The transmitted signal is then   collected and analyzed  by the receiver detection module.
	The connection between  sent and received quantum states can be established  with the  input-output relation for the quantum state~\cite{Semenov2009}
		\begin{align}\label{Pinout}
		P_{\rm out}(\alpha)=\int_0^1\D\eta\mathcal{P}(\eta)\frac{1}{\eta}P_{\rm in}\left(\frac{\alpha}{\sqrt{\eta}}\right).
		\end{align}
	This relation is written in terms of the Glauber-Sudarshan  quasiprobability distributions \cite{Glauber, Sudarshan} $P_{\mathrm{in}}(\alpha)$ and $P_{\mathrm{out}}(\alpha)$ of the input and output quantum states, respectively.
	Here $\mathcal{P}(\eta)$ is the PDT for the  atmospheric channel and $\eta\in[0,1]$ is the intensity transmittance.
	Provided the PDT is known for a specific quantum channel, the analysis of quantum properties of the transmitted state  can be performed straightforwardly from Eq.~(\ref{Pinout}).
	An important fact is that the PDT is a positive-semidefinite function in both quantum and classical theories and can be obtained from purely classical models or experiments.

	In many practical situations the detection scheme on the receiver site has  a telescope or another focusing unit that collects the transmitted signal for further detection.
	Since the telescope has  a finite entrance  pupil the transmitted signal arriving on the detector is influenced  by the finiteness of the entrance aperture.
	This action of the receiver aperture is superimposed with the random distortions of the optical signal caused by the turbulent atmosphere and results in a fluctuating transmittance which can be written as
		\begin{align}\label{eta}
		\eta=\int\limits_{\mathcal{A}}\D^2\boldsymbol{r} I(\boldsymbol{r},L).
		\end{align}
	Here  $I(\boldsymbol{r},L)$ is the normalized intensity of a classical beam, $L$ is the beam propagation distance, and $\mathcal{A}$ is the area of the receiver aperture, which we assume to be circular with the radius $a$.
	In Eq.~(\ref{eta}) the transverse spatial coordinate $\boldsymbol{r}$ is chosen in such a way that $\boldsymbol{r}{=}0$ coincides with the center of the aperture opening.
	The transmittance is a randomly varying quantity with values $\eta\in[0,1]$; its statistical properties for the specific atmospheric quantum channel are given by the corresponding PDT.

	Among the most pronounced effects that influence the value of atmospheric transmittance (\ref{eta}) are  random deflections of the light beam as a whole by turbulent inhomogeneities (beam wandering), turbulence-induced beam broadening, and deformation.
	The PDT that accounts for the beam-wandering effect was derived in Ref.~\cite{Vasylyev2012} and was further extended in Ref.~\cite{Vasylyev2016} in order to include the effects of beam broadening and deformation into an elliptic form.
	The elliptic-beam model allows one to obtain a consistent PDT that agrees well with experimental data \cite{Vasylyev2017}.	
	It requires elaborate calculations of statistical parameters.
	The difficulty of this task grows essentially in the regime of moderate turbulence~\cite{Tatarskii1979, Leader, Chumak, Chumak1}.
	The given form of the elliptic-beam approximation assumes special statistics for the shape characteristics of the transmitted beam.
	Although this model shows proper results for relatively short propagation distances, the statistical assumptions should be reconsidered for long-distance channels.
	
	In this paper we introduce an alternative way to overcome this problem.
	The main idea of this approach consists in the separation of contributions from beam wandering and beam-spot distortions by applying the law of total probability \cite{Schervish}.
	The resulting PDT describes practically all atmospheric channels with initially Gaussian beams.
	It depends on the proper knowledge of the classical field-correlation functions of the second and fourth orders.
	A problem is that calculations of the field-correlation functions require applications of involved numerical methods, which do not work properly in all cases.
	For this reason, we propose an approximation, which assumes relatively weak contributions from beam wandering.
	The resulting PDT depends only on four parameters: the first two moments of the transmittance $\langle\eta\rangle$ and $\langle\eta^2\rangle$, the beam-wandering variance  $\sigma_{\mathrm{bw}}$, and the short-term radius  $W_\mathrm{ST}$, of the beam spot.
	These parameters are related to field-correlation functions of the second and fourth orders.
	Nevertheless, their determination requires applications of fewer computational resources.
	Moreover, the separation of the contribution from beam wandering allows one to derive the PDT for the case when the  beam-tracking procedure is applied.

	The paper is organized as follows. In Sec.~\ref{sec:Preliminaries} we consider two known PDT models  corresponding to two limiting cases of vanishing and dominant contribution of beam wandering.
	In Sec.~\ref{sec:TheoreticalModel} we introduce the  method for the calculation of the PDT based on  the law of total probability.
	The calculation of the PDT requires the knowledge of the conditional moments of the atmospheric transmittance.
	In Sec.~\ref{sec:Approx} we introduce approximative formulas for these quantities.
	The developed theory is applied then to the description of atmospheric quantum channels in Sec.~\ref{sec:Applications}.
	In Sec.~\ref{sec:Tracking} we extend the PDT model in order to describe the beam-tracking procedure.
	In Sec.~\ref{sec:decoy} the PDT theory is applied to the security analysis of the two-decoy state quantum protocol.
	A summary and some conclusions are given in Sec.~\ref{sec:Summary}.


\section{Preliminaries}\label{sec:Preliminaries}
\subsection{Field correlation functions}

	In the framework of classical atmospheric optics, for a complete description of the propagation of optical radiation in the turbulent media it is necessary to know the probability distribution functional of the random radiation field.
	Since its determination is an extremely complicated problem, the knowledge of the first correlation functions is usually used for the characterization of classical atmospheric optical channels.
	The second-order field-correlation function
		\begin{align}\label{Gamma2}
		\Gamma_2(\boldsymbol{r},L)=\langle I(\boldsymbol{r},L)\rangle
		\end{align}
	serves as the mean intensity of the radiation scattered in a randomly inhomogeneous medium.
	It is also used for the  determination of the beam spreading caused by atmospheric turbulence and for the characterization of  the  spatial coherence of the beam.
	The fourth-order field-correlation function
		\begin{align}\label{Gamma4}
		&\Gamma_4(\boldsymbol{r}_1,\boldsymbol{r}_2,L){=}\langle I(\boldsymbol{r}_1,L)I(\boldsymbol{r}_2,L)\rangle
		\end{align}
	describes the intensity fluctuations of optical radiation.
	The correlation function $\Gamma_4$ is crucial for the examination of fourth-order statistical quantities such as the scintillation index, the irradiance covariance function, the beam-wandering variance, etc.

	The field-correlation functions $\Gamma_2$ and $\Gamma_4$ play an important role in the description of atmospheric  quantum channels.
	Using the definition (\ref{eta}), the first two moments  of the  channel transmittance or, alternatively, the two moments of the PDT, are related to the field-correlation functions as
		\begin{align}\label{MEta}
		\langle\eta\rangle=\int\limits_{\mathcal{A}}\D^2\boldsymbol{r} \Gamma_2(\boldsymbol{r},L),
		\end{align}
		\begin{align}\label{MEta2}
		\langle\eta^2\rangle=\int\limits_{\mathcal{A}}\D^2\boldsymbol{r}_1 \int\limits_{\mathcal{A}}\D^2\boldsymbol{r}_2 \Gamma_4(\boldsymbol{r}_1,\boldsymbol{r}_2,L).
		\end{align}
	Using Eqs.~(\ref{MEta}) and (\ref{MEta2}), one can also calculate the transmittance variance,  $\langle(\Delta\eta)^2\rangle{=} \langle\eta^2\rangle- \langle\eta\rangle^2$, which is an important characteristics of the PDT.
	The parameter $\langle(\Delta\eta)^2\rangle/\langle\eta\rangle^2$, on the other hand, is used in classical atmospheric optics to account for the aperture averaging of the scintillation index~\cite{Tatarskii, Yura1983}.

\subsection{Truncated log-normal distribution}\label{sec:truncated}

	The log-normal distribution is widely used in classical and quantum atmospheric optics \cite{Diament1970,Perina1972,Perina1973,Milonni, Tatarskii, Bohmann2017}.
	However, this distribution was originally applied to the description of the random-beam intensity in a given spatial point, $I(0)$.
	In the context of the present consideration, this value reads 
		\begin{equation}
		 I(0)=\lim_{\mathcal{A}\rightarrow 0} \frac{\eta(\mathcal{A})}{\mathcal{A}},
		\end{equation}
	where $\eta(\mathcal{A})$ is a function of the aperture area $\mathcal{A}$ as given by Eq. \eqref{eta}.
	Unlike the efficiency $\eta$, the intensity $I(0)$ can attain arbitrary high values.
	Hence, the log-normal distribution for $I(0)\in[0,+\infty)$   can be determined consistently.
	On the other hand, the transmittance $\eta$ for a finite aperture area $\mathcal{A}$ must be restricted to the domain $[0,1]$.
	For this reason, the log-normal distribution for $\eta$, in the cases of appropriate propagation scenarios, must be vanishing at the value of $\eta{=}1$.

	For some cases with long propagation lengths or strong turbulence, the effects of beam-spot distortions significantly dominate  the resulting statistics, compared to the effects of beam wandering.
	In this case, the PDT can be approximated with  reasonable accuracy by the truncated log-normal distribution (cf. Ref. \cite{Capraro} for an experiment and Ref. \cite{Vasylyev2016} for a theoretical explanation)
		\begin{align}\label{lnTrunc}
		\mathcal{P}(\eta)&= \mathcal{P}(\eta;\mu,\sigma)=\nonumber\\
		&=\left\{\begin{array}{l c}
		\frac{1}{\mathcal{F}(1)}\frac{1}{\sqrt{2\pi}\eta\sigma}\exp\Bigl[-\frac{(\ln \eta+\mu )^2}{2\sigma^2}\Bigr]&\text{for}\,\eta\in[0,1]\\
		0&\text{otherwise,}
			\end{array}\right. 
		\end{align}
	where  $\mathcal{F}(1)$ is the cumulative function of the (nontruncated) log-normal distribution at the point $\eta{=}1$.
	The parameters $\mu$ and $\sigma$ in Eq.~(\ref{lnTrunc}) can be approximately expressed through the transmittance moments (\ref{MEta}) and (\ref{MEta2}) as
		\begin{align}\label{Mu}
		\mu=\mu(\langle\eta\rangle,\langle\eta^2\rangle)\approx-\ln\left[\frac{\langle\eta\rangle^2}{\sqrt{\langle\eta^2\rangle}}\right],
		\end{align}
		\begin{align}\label{Sigma}
		\sigma=\sigma(\langle\eta\rangle,\langle\eta^2\rangle)\approx\sqrt{\ln\left[\frac{\langle\eta^2\rangle}{\langle\eta\rangle^2}\right]}.
		\end{align}
	Here and in the following the approximation sign is used to signify that the exact expressions for the truncated log-normal distribution are replaced with the corresponding expressions for the standard log-normal distribution.
	In the case when the  value  $\mathcal{P}(1;\mu,\sigma)$ is vanishingly small, the expressions (\ref{Mu}) and (\ref{Sigma}) become almost exact.

	It is also important that the parameters of the truncated log-normal distribution should be calculated by using Eqs. \eqref{MEta}, \eqref{MEta2}, \eqref{Mu}, and \eqref{Sigma}.
	These equations include the finite size of the aperture.
	An incorrect usage of these rules and improper truncations may result in unphysical effects such as the fake creation of photons by atmospheric turbulence \cite{Perina1973}.

\subsection{Beam wandering distribution}

	For short propagation distances and weak turbulence, the contribution of beam-wandering effects in the resulting statistics has a dominating character.
	In this case, the PDT significantly differs from the truncated log-normal distribution and has the form of the log-negative Weibull distribution \cite{Vasylyev2012}.
	This model assumes that the fluctuating losses of the transmitted optical beam originate from the random deflections of the beam centroid on turbulent inhomogeneities.
	Furthermore, we assume that the beam profile at the aperture plane can be approximated by a Gaussian shape.
	
	The random transverse vector $\boldsymbol{r}_0$, which describes the position of the deflected beam centroid relative to the aperture center, is considered to be normally distributed.
	The corresponding probability distribution is given by
			\begin{align}\label{BWprob}
			\rho(\boldsymbol{r}_0)=\frac{1}{2\pi\sigma_{\mathrm{bw}}^2}\exp\left[-\frac{\boldsymbol{r}_0^2}{2\sigma_{\mathrm{bw}}^2}\right],
			\end{align}
	where
			\begin{align}\label{SigmaBW}
			\sigma_{\mathrm{bw}}^2=\int_{\mathbb{R}^4}\D\mathbf{r}_1\D\mathbf{r}_2 x_1 x_2\Gamma_4(\mathbf{r}_1,\mathbf{r}_2,L)
			\end{align}
	is the beam-wandering variance \cite{Kon}.
	The beam has the spot width $W_{\mathrm{ST}}$ given by
			\begin{align}\label{Wst}
			W_{\mathrm{ST}}^2=4\Bigl[\int_{\mathbb{R}^2}\D\mathbf{r} x^2\Gamma_2(\mathbf{r},L)- \sigma_{\mathrm{bw}}^2\Bigr],
			\end{align}
	which is the short-term beam broadening \cite{Fante}.

	The transmittance of the Gaussian beam deflected by the distance $r_0=|\boldsymbol{r}_0|$ from the aperture center can be analytically approximated as
		\begin{align}\label{bw_transmittance}
		\eta=\eta_0\exp\left[-\left(\frac{r_0}{R}\right)^{\lambda}\right].
		\end{align}
	The parameters  $\eta_0$, $R$, and $\lambda$  (cf. Appendix~\ref{app:parameters}) are  functions of the aperture radius $a$ and the short-term width $W_{\mathrm{ST}}$.
	This yields the analytical form of the beam-wandering PDT \cite{Vasylyev2012}
		\begin{align}\label{Pbw}
		\mathcal{P}(\eta)&=\frac{R^2}{\sigma_{\mathrm{bw}}^2\eta\,\lambda}\left(\ln\frac{\eta_0}{\eta}\right)^{\frac{2}{\lambda}-1}\nonumber\\
		&\times
		\exp\left[-\frac{R^2}{2\sigma^2_{\mathrm{bw}}}\left(\ln\frac{\eta_0}{\eta}\right)^{(2/\lambda)}\right],
		\end{align}
	for $\eta\in[0,\eta_0]$ and $\mathcal{P}(\eta)=0$ otherwise, which is the log-negative Weibull distribution.


\section{Theoretical model}\label{sec:TheoreticalModel}

	The truncated log-normal distribution and the beam-wandering model are two limiting cases that do not describe all possible situations of light propagation in the turbulent atmosphere.
	In Ref.~\cite{Vasylyev2016}, the so-called elliptic-beam model was introduced in order to describe other effects beyond beam wandering.
	The elliptic-beam model adequately describes the regime of weak-to-moderate turbulence; it shows a behavior similar to the truncated log-normal  distribution for the regime of strong  turbulence.
	However, the given form of the elliptic-beam approximation does not  work  in some important cases, e.g., for long-distance propagation in atmospheric channels.
	In such cases the first two moments of the transmittance, $\langle\eta\rangle$ and $\langle\eta^2\rangle$, obtained from the elliptic-beam model, may significantly differ from those obtained from Eqs. \eqref{MEta} and  \eqref{MEta2}.

\begin{figure}[ht]
		\includegraphics[width=0.32\textwidth]{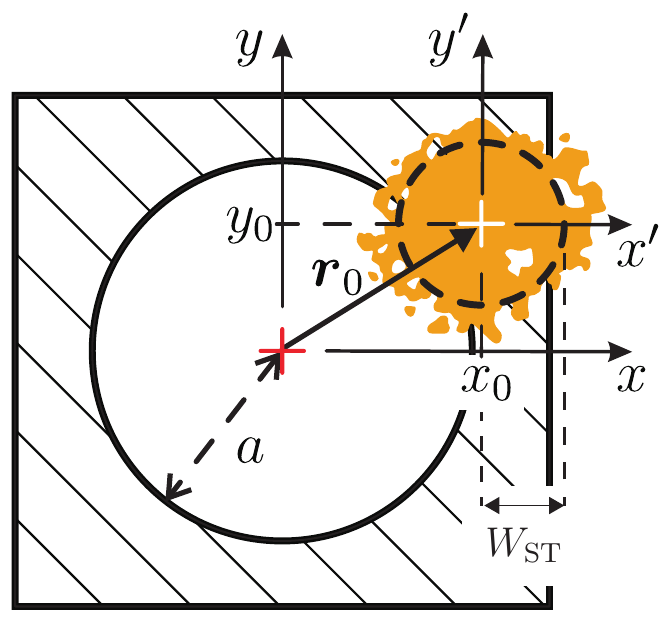}
		\caption{\label{fig:Beam} Transmitted beam profile shown relative to the aperture opening of radius $a$.
		The beam centroid is displaced relative to the  aperture center due to beam wandering and its position is given by $\boldsymbol{r}_0$.
		The tracking coordinate system $(x^\prime,y^\prime)$ is connected with the fluctuating position of the beam centroid.	}
		\end{figure}

	The propagation of laser radiation over long distances in the atmospheric turbulence is followed by a decrease of coherence, wavefront distortion, and fluctuations of  beam amplitude and phase.
	The intensity profile  of the transmitted beam has an irregular form and is randomly displaced from the receiver aperture opening due to beam wandering (cf.~Fig.~\ref{fig:Beam}).
	The fluctuations of the aperture transmittance (\ref{eta}) evidently depend on the instantaneous beam profile and the position of its centroid relative to the aperture center.
	This observation facilitates the derivation of the PDT for general situations where the transmitted beam profile could have an arbitrary form.

	In the present paper, we propose to derive the PDT  based on the idea of splitting the contributions from beam wandering and beam-spot distortion effects.
	Using the law of total probability \cite{Schervish}, the PDT can be written  as
	\begin{align}\label{Bayesian}
	\mathcal{P}(\eta)=\int_{\mathbb{R}^2}  \D^2 \boldsymbol{r}_0 P(\eta|\boldsymbol{r}_0)\rho(\boldsymbol{r}_0).
	\end{align}	
	The beam-wandering contribution is described by the distribution $\rho(\boldsymbol{r}_0)$, which is given by Eq.~(\ref{BWprob}) and depends on the beam-wandering variance (\ref{SigmaBW}).
	The effects of beam-spot distortions are incorporated in the conditional probability $P(\eta|\boldsymbol{r}_0)$.
	This function can be interpreted as the conditional PDT if the beam would be tracked to the position $\boldsymbol{r}_0$  relative to the aperture center.

	In fact, Eq. \eqref{Bayesian} resembles the method of elliptic-beam approximation \cite{Vasylyev2016}.
	In this case, $P(\eta|\boldsymbol{r}_0)$ is the distribution obtained with the assumption that after passing the atmosphere the beam has the form of a random ellipse under the condition that it is deflected by the distance $r_0$ from the aperture center.
	As it has been discussed in Ref. \cite{Vasylyev2016}, this function is very well approximated by the truncated log-normal distribution, such that
		\begin{align}\label{Pcond}
		P(\eta|\boldsymbol{r}_0)\approx\mathcal{P}(\eta;\mu_{r_0},\sigma_{r_0}).
		\end{align}	
	The parameters of this distribution are  expressed through the conditional transmittance moments as
		\begin{align}\label{mur}
		\mu_{r_0}\approx\mu(\langle\eta\rangle_{r_0},\langle\eta^2\rangle_{r_0}),\qquad  \sigma_{r_0}\approx\sigma(\langle\eta\rangle_{r_0},\langle\eta^2\rangle_{r_0}),
		\end{align}
	with the explicit approximate dependence given by Eqs.~(\ref{Mu}) and (\ref{Sigma}).
	The usage of the truncated log-normal distribution leaves an open question about tiny details of tail behavior for $\eta{\approx} 1$ (cf.~Ref.~\cite{Vasylyev2016}).

	The parameters $\mu_{r_0}$ and $\sigma_{r_0}$ uniquely define the truncated log-normal distribution \eqref{Pcond}.
	This distribution in turn is used in the law of total probability \eqref{Bayesian} for obtaining the PDT.
	In the following we describe the technique of obtaining these parameters from the field-correlation functions of the second and fourth orders $\Gamma_2$ and $\Gamma_4$, respectively.
		
	Let us consider the aperture plane and the optical beam impinging on the aperture (see Fig.~\ref{fig:Beam}).
	We denote by $I^{(c)}(\boldsymbol{r}^\prime,L)$  the intensity of the beam in the coordinates $\boldsymbol{r}^\prime=(x^\prime,y^\prime)$.
	The origin of this coordinate system is not fixed and coincides with the fluctuating position of the beam centroid  such that $I^{(c)}(\boldsymbol{r}^\prime,L)$ can be considered as the perfectly-beam-tracked intensity.
	This intensity has the  obvious connection to the intensity $I(\boldsymbol{r},L)$ from Eq.~(\ref{eta}), which reads 
		\begin{align}
		I(\boldsymbol{r},L)=I^{(c)}(\boldsymbol{r}-\boldsymbol{r}_0,L)
		\end{align}
	and is obtained with the help of the coordinate transformation $\boldsymbol{r}^\prime=\boldsymbol{r}-\boldsymbol{r}_0$ (cf.~Fig.~\ref{fig:Beam}).
	Assuming that the beam centroid deflection $\boldsymbol{r}_0$ is normally distributed according to the probability distribution $\rho(\boldsymbol{r}_0)$ [cf. Eq.~(\ref{BWprob})] and using the  correlation function (\ref{Gamma2}), we obtain
		\begin{align} \label{ConditGamma2}
		\Gamma_2(\boldsymbol{r},L)=\int_{\mathbb{R}^2} \D^2\boldsymbol{r}_0  \rho(\boldsymbol{r}_0)\Gamma_{2}^{(c)}(\boldsymbol{r}-\boldsymbol{r}_0,L),
		\end{align}
	where $\Gamma_2^{(c)}(\boldsymbol{r},L)=\langle I^{(c)}(\boldsymbol{r},L)\rangle$ is the second-order correlation function of the perfectly-tracked beam.
	Similarly, from Eq.~(\ref{Gamma4}) we derive
		\begin{align}\label{ConditGamma4}
		\Gamma_4(\boldsymbol{R},\boldsymbol{\rho},L){=}\int_{\mathbb{R}^2} \D^2\boldsymbol{r}_0  \rho(\boldsymbol{r}_0)\,
		\Gamma_{4}^{(c)}(\boldsymbol{R}{-}\sqrt{2}\boldsymbol{r}_0,\boldsymbol{\rho},L),
		\end{align}
	where the coordinates, $ \boldsymbol{R}{=}(\boldsymbol{r}_1{+}\boldsymbol{r}_2)/\sqrt{2}$ and $\boldsymbol{\rho}{=}(\boldsymbol{r}_1{-}\boldsymbol{r}_2)/\sqrt{2}$, are used.

	Equation~(\ref{ConditGamma2}) can be inverted with respect to $\Gamma_{2}^{(c)}$ by using the inverse Weierstrass transform \cite{Brychkov}
		\begin{align}\label{DiffGamma2}
		\Gamma_{2}^{(c)}(\boldsymbol{r},L)=\exp\left[-\frac{\sigma_{\mathrm{bw}}^2}{2}\Delta_{\boldsymbol{r}}\right]\Gamma_2(\boldsymbol{r},L),
		\end{align}
	where $\Delta_{\boldsymbol{r}}{=}\frac{\partial^2}{\partial x^2}{+}\frac{\partial^2}{\partial y^2}$ is the transverse Laplace operator.
	Similarly, one can invert Eq.~(\ref{ConditGamma4}) with respect to $\Gamma_{4}^{(c)}$ by performing the inverse Weierstrass transform,
		\begin{align}\label{DiffGamma4}
		\Gamma_{4}^{(c)}(\boldsymbol{R},\boldsymbol{\rho},L)=\exp\left[-\sigma_{\mathrm{bw}}^2\Delta_{\boldsymbol{R}}\right]
		\Gamma_{4}(\boldsymbol{R},\boldsymbol{\rho},L).
		\end{align}
	This means that the functions $\Gamma_{2}^{(c)}(\boldsymbol{r},L)$ and $\Gamma_{4}^{(c)}(\boldsymbol{R},\boldsymbol{\rho},L)$ can be obtained as solutions of  heat   equations with negative diffusion coefficients and with the initial conditions given by the functions $\Gamma_{2}(\boldsymbol{r},L)$ and $\Gamma_{4}(\boldsymbol{R},\boldsymbol{\rho},L)$, respectively.
	The obtained functions $ \Gamma_{2}^{(c)}$ and $ \Gamma_{4}^{(c)}$ can be used for calculating the conditional moments $\langle\eta\rangle_{r_0}$ and $\langle\eta^2\rangle_{r_0}$ in analogy to Eqs.~(\ref{MEta}) and (\ref{MEta2}),
		\begin{align}\label{MEtar}
		\langle\eta\rangle_{r_0}=\int\limits_{\mathcal{A}(\boldsymbol{r}_0)}\D^2\boldsymbol{r} \Gamma_2^{(c)}(\boldsymbol{r},L),
		\end{align}
		\begin{align}\label{MEta2r}
		\langle\eta^2\rangle_{r_0}=\int\limits_{\mathcal{A}(\boldsymbol{r}_0)}\D^2\boldsymbol{r}_1
		\int\limits_{\mathcal{A}(\boldsymbol{r}_0)}\D^2\boldsymbol{r}_2 \Gamma_4^{(c)}(\boldsymbol{r}_1,\boldsymbol{r}_2,L),
		\end{align}
	where $\mathcal{A}({\boldsymbol{r}}_0)$ denotes the circular aperture opening, its center  being displaced relative to the beam centroid position by $\boldsymbol{r}_0$.

	The explicit dependence of the conditional moments (\ref{MEtar}) and (\ref{MEta2r}) on the displacement parameter can be obtained, provided the field-correlation functions $\Gamma_2$ and $\Gamma_4$ in Eqs.~(\ref{DiffGamma2}) and (\ref{DiffGamma4}) are known.
	These parameters uniquely define the conditional probability distribution $P(\eta|\boldsymbol{r}_0)$, which is used in the law of total probability \eqref{Bayesian} for obtaining the PDT.
	In practice this approach requires an application of involved numerical methods.
	For this reason, we propose a method, which enables one to overcome this problem with fewer computational resources.


\section{Approximation of weak beam wandering}\label{sec:Approx}

	For the PDT model considered here, the first two moments of the transmittance $\langle\eta\rangle$ and $\langle\eta^2\rangle$ exactly agree with the values obtained from Eqs. \eqref{MEta} and \eqref{MEta2}.
	We aim to preserve this important property by developing an approximation method for the calculation of the integrals in Eqs. \eqref{MEtar} and \eqref{MEta2r}, which give the conditional moments $\langle\eta\rangle_{r_0}$ and $\langle\eta^2\rangle_{r_0}$.
	As a consequence of the law of total probability \eqref{Bayesian}, the conditional moments are related to the moments of the transmittance $\eta$ via
		\begin{align}\label{moment1}
		\langle\eta\rangle=\int_{\mathbb{R}^2}  \D^2 \boldsymbol{r}_0 \rho(\boldsymbol{r}_0)\langle\eta\rangle_{r_0},
		\end{align}
		\begin{align}\label{moment2}
		\langle\eta^2\rangle=\int_{\mathbb{R}^2}  \D^2 \boldsymbol{r}_0 \rho(\boldsymbol{r}_0)\langle\eta^2\rangle_{{r}_0},
		\end{align}
	where $\rho(\boldsymbol{r}_0)$ is given by Eq.~(\ref{BWprob}).
	
	An explicit dependence of the first conditional moment $\langle\eta\rangle_{r_0}$ on $r_0$ can be obtained from the following physical considerations.
	Taking into account that the field-correlation functions   $\Gamma_{2}^{(c)}(\boldsymbol{r},L)$ and  $\Gamma_{2}(\boldsymbol{r},L)$ have Gaussian forms  to a good approximation~\cite{Brown, Yura1972},  $\Gamma_{2}^{(c)}(\boldsymbol{r},L)$ can be interpreted as the intensity of an effective perfectly tracked Gaussian beam 
	with the beam width given by the short beam-spot width $W_\mathrm{ST}$.
	Consequently, the conditional moment $\langle\eta\rangle_{r_0}$ can be considered as the transmittance through the aperture of this effective beam at the distance $r_0$ from the aperture center [cf. Eq. \eqref{bw_transmittance}].
	This yields
		\begin{align}\label{CondEta1}
		\langle\eta\rangle_{r_0}=\eta_0\exp\left[-\left(\frac{r_0}{R}\right)^{\lambda}\right],
		\end{align}
	where $\eta_0$, $R$, and $\lambda$ are expressed by the aperture radius $a$ and the short-term width $W_{\mathrm{ST}}$  [cf. Appendix~\ref{app:parameters}].
	For the case of Gaussian $\Gamma_{2}(\boldsymbol{r},L)$, the $\langle\eta\rangle$ obtained from Eqs. \eqref{moment1} and \eqref{MEta} coincide.
	If $\Gamma_{2}(\boldsymbol{r},L)$ significantly deviates from the Gaussian form, the parameter $\eta_0$ should be specified as
		\begin{align}\label{Eta1Defin}
		\eta_0=\frac{\langle\eta\rangle}{\displaystyle{\int\limits_{0}^\infty  \D\xi\,\xi \,e^{-\frac{\xi^2}{2}}e^{-\left(\frac{\sigma_{\mathrm{bw}}}{R}\xi\right)^{\lambda}}}}.
		\end{align}
	This equation is derived via substituting Eq. \eqref{CondEta1} into Eq. \eqref{moment1} and then expressing $\eta_0$ explicitly.
		
	A similar consideration for the second conditional moment $\langle\eta^2\rangle_{r_0}$ requires additional assumptions.
	In order to formulate them, we note that some obvious restrictions should be satisfied:
        (i) The conditional variance is a non-negative function  $\langle(\Delta\eta)^2\rangle_{r_0}=\langle\eta^2\rangle_{r_0}-\langle\eta\rangle_{r_0}^2{\ge}0$ and
	the  conditional  exceedance
		\begin{align}
		\overline{\mathcal{F}}(\eta|\boldsymbol{r}_0)=\int_\eta^1\D\eta^\prime P(\eta^\prime|\boldsymbol{r}_0),
		\end{align}
	i.e., the  probability that the transmittance exceeds the value $\eta$ under the condition that the beam centroid is displaced from the aperture center by $\boldsymbol{r}_0$, obeys the inequality
		\begin{align}\label{ExceedInequality}
		\overline{\mathcal{F}}(\eta|\boldsymbol{r}_0)\ge \overline{\mathcal{F}}(\eta|\boldsymbol{r}_0^\prime),\quad\text{for}\quad {r}_0\le {r}_0^\prime,
		\end{align}
	which  means that increasing the beam displacement cannot improve the transmission characteristics.

	An approximation for the second conditional moment $\langle\eta^2\rangle_{{r}_0}$, which satisfies these requirements, can be obtained by assuming small values of the beam-wandering variance $\sigma_{\mathrm{bw}}^2$.
	Let us consider the conditional aperture-averaged scintillation index as a function of the beam deflection $r_0$,
		\begin{align}\label{sint}
 		\sigma^2_\textrm{sc}(r_0)=\frac{\langle\Delta\eta^2\rangle_{r_0}}{\langle\eta\rangle_{r_0}^2}=\frac{\langle\eta^2\rangle_{r_0}-\langle\eta\rangle_{r_0}^2}{\langle\eta\rangle_{r_0}^2}.
 		\end{align}
	For the case of small values $r_0$, i.e., for weak beam wandering, $\langle\eta\rangle_{r_0}$ according to Eq. \eqref{CondEta1} is a slightly varying function of $r_0$.
	The same behavior can be assumed for $\langle\eta^2\rangle_{r_0}$.
	Consequently, by expanding $\sigma^2(r_0)$ in a Taylor series with respect to $r_0$, we can restrict ourselves to the zeroth-order term only, i.e., we assume that $\sigma^2_\textrm{sc.}(r_0)=\textrm{const}$.
	In fact, this means that in the region of such displacements of the tracked beam relative to the aperture only lead  to additional deterministic losses.\footnote{The scintillation index does not depend on the deterministic (nonfluctuating) losses.  }
	One can provide additional arguments  supporting this approximation.
	First, there is the observation \cite{Churnside, Andrews1999} that the aperture-averaged scintillation index is almost independent on the beam area that passes through the aperture in a wide domain of its values.
	Second, the experiments in Refs. \cite{Vorontsov, Gurvich} demonstrate a weak dependence of the scintillation index on the position of  the observation point.

	This assumption of weak beam wandering  yields
		\begin{align}\label{CondEta2}
		\langle\eta^2\rangle_{r_0}=\zeta_{0}^2\exp\left[-2\left(\frac{r_0}{R}\right)^{\lambda}\right].
		\end{align}
	The parameter  $\zeta_{0}^2$  is determined by substituting Eq. (\ref{CondEta2}) into Eq. (\ref{moment2}) and reads as
		\begin{align}\label{Eta2Defin}
		\zeta_0^2=\frac{\langle\eta^2\rangle}{\displaystyle{\int\limits_{0}^\infty  \D\xi\,\xi \,e^{-\frac{\xi^2}{2}}e^{-2\left(\frac{\sigma_{\mathrm{bw}}}{R}\xi\right)^{\lambda}}}}.
		\end{align}
	Inserting Eqs.~(\ref{CondEta1}) and (\ref{CondEta2}) into (\ref{mur}),  we obtain explicit expressions for the parameters of the truncated log-normal distribution (\ref{Pcond}),
		\begin{align}\label{muApprox}
		\mu_{r_0}\approx-\ln\left[\frac{\eta_0^2}{\zeta_0}\right]+\left(\frac{r_0}{R}\right)^{\lambda},
		\end{align}
		\begin{align}\label{sigmaApprox}
		\sigma_{r_0}\approx\sqrt{\ln\left[\frac{\zeta_0^2}{\eta_0^2}\right]}.
		\end{align}
	The assumption that the aperture-averaged scintillation index,  $\sigma^2_\textrm{sc.}(r_0)$ does not depend on the displacement $r_0$ yields a constant value of the log-normal parameter $ \sigma_{r_0}$.
	The obtained parameters $\mu_{r_0}$ and $\sigma_{r_0}$ uniquely define the conditional PDT $P(\eta|\boldsymbol{r}_0)$ [cf. Eq. \eqref{Pcond}] which is used in the law of total probability \eqref{Bayesian} for determining the   PDT of the channel under study.
	The corresponding step-by-step procedure is summarized in Appendix~\ref{app:PDTeval}.
	
	The obtained PDT describes the discussed variations of the PDT, depending on the channel characteristics, between log-negative Weibull and truncated log-normal distributions in a mathematically correct way.
	However, some details of the PDT may have significant errors for cases of increasing beam-wandering variance.
	This result should be considered as an approximation and further improvements of the model may be in order.


\section{Application to atmospheric channels}\label{sec:Applications}

	We illustrate the proposed approach for the PDT calculations by considering several atmospheric channels with diverse propagation conditions.
	For this purpose we choose the short-distance atmospheric links of $L{=}$~1, 2, and 3~km length.
	We calculate the field correlation functions (\ref{Gamma2}) and (\ref{Gamma4}) of the transmitted light by using the phase approximation of the  Huygens-Kirchhoff method (cf.~\cite{Banakh}  and the Supplemental Matherial of Ref.~\cite{Vasylyev2016})\footnote{
	It is noteworthy that the phase approximation does not take into account amplitude fluctuations which mostly contribute in the regime of moderate turbulence~\cite{Holmes}.
	For  calculations in this regime, appropriate methods are the extended Huygens-Fresnel method~\cite{Holmes} or the photon distribution function approach \cite{Chumak, Chumak1}.  }.
	In the first order of this approximation one derives for the focused beam
		\begin{align}\label{Gamma2Calc}
		\Gamma_2(\mathbf{r},L)=\frac{k^2}{4\pi^2L^2}\int_{\mathbb{R}^3}\D^2\mathbf{r}^\prime\,e^{-\frac{|\mathbf{r}^\prime|^2}{2W_0^2}
		-2i\frac{\Omega}{W_0^2}\mathbf{r}{\cdot}\mathbf{r}^\prime{-}\frac{1}{2}\mathcal{D}_S(0,\mathbf{r}^\prime)},
		\end{align}
		\begin{align}\label{Gamma4Calc}
		&\Gamma_4(\mathbf{r}_1,\mathbf{r}_2,L)=\frac{2k^4}{\pi^2(2\pi)^3L^4W_0^2}\int_{\mathbb{R}^6}\D^2\mathbf{r}_1^\prime
		\D^2\mathbf{r}_2^\prime\D^2\mathbf{r}_3^\prime\nonumber\\
		&\quad\times e^{-\sum\limits_{i=1}^3\frac{|\mathbf{r}_i^\prime|^2}{W_0^2}-2i\frac{\Omega}{W_0^2}\left[(\mathbf{r}_1
		{-}\mathbf{r}_2){\cdot}\mathbf{r}_2^\prime{+}(\mathbf{r}_1{+}\mathbf{r}_2){\cdot}\mathbf{r}_3^\prime\right]}\\
		&\quad\times\exp\Bigl[\frac{1}{2}\sum\limits_{j=1,2}\Bigl\{\mathcal{D}_S(\mathbf{r}_1{-}\mathbf{r}_2,
		\mathbf{r}_1^\prime{+}(-1)^j\mathbf{r}_2^\prime)\nonumber\\
		&-\mathcal{D}_S(\mathbf{r}_1{-}\mathbf{r}_2,\mathbf{r}_1^\prime{+}(-1)^j\mathbf{r}_3^\prime)-
		\mathcal{D}_S(0,\mathbf{r}_2^\prime{+}(-1)^j\mathbf{r}_3^\prime)\Bigr\}\Bigr],\nonumber
		\end{align}
	where for the  Kolmogorov-Obukhov turbulence spectrum   \cite{Tatarskii} the phase structure function reads 
		\begin{align}
		&\mathcal{D}_S(\mathbf{r},\mathbf{r}^\prime)=2C_n^2 k^2 L\int_0^1\D\xi\left|\mathbf{r}\xi+\mathbf{r}^\prime(1-\xi)\right|^{\frac{5}{3}}.
		\end{align}
	Here $k$ is the optical wave number,  $C_n^2$ is the turbulence refractive-index structure constant, $L$ is the propagation length, $W_0$ is the beam-spot width at the transmitter site, and $\Omega{=}kW_0^2/2L$ is the Fresnel parameter.
	In general, the calculation of the correlation functions and their moments requires high-accuracy numerical integrations~\cite{Chumak, Kravtsov, Jakeman}.

	The level of optical-beam distortion  in the turbulent atmosphere depends on the parameters of the atmosphere, on the optical channel length, and on the  wavelength of the optical signal.
	The (local) strength of turbulence in the atmosphere is determined by the refractive index structure constant $C_n^2$.
	Under diverse daytime and weather conditions, geographical location, and altitude,  its value  varies typically from $10^{-17}\text{m}^{-3/2}$ (weak turbulent fluctuations) to $10^{-12}\text{m}^{-3/2}$ (strong fluctuations).
	The strength of optical turbulence  incorporates besides $C_n^2$ also the propagation length  and   optical-beam wave number and is usually associated with  the Rytov parameter
		\begin{align}\label{RytovSigma}
		\sigma_R^2 =1.23 C_n^2 k^{\frac{7}{6}}L^{\frac{11}{6}}.
		\end{align}
	Depending on the value of the Rytov parameter we distinguish weak ($\sigma_R^2< 1$), weak to moderate ($\sigma_R^2\approx 1$), moderate ($\sigma_R^2>1$), and strong ($\sigma_R^2\gg 1$) optical turbulence.
	This classification is merely related to the strength of intensity fluctuations of the transmitted light~\cite{Kravtsov}.

\begin{figure}[ht]
 \includegraphics[width=0.45\textwidth]{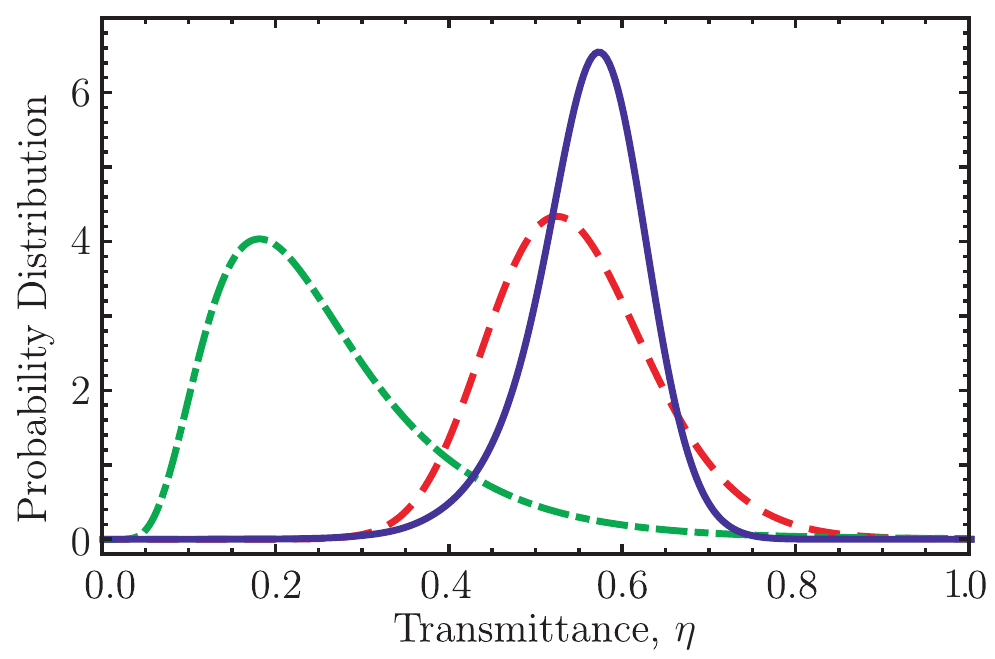}
 \caption{\label{fig:Bayes1}  The PDTs for the light beam  transmitted through the atmospheric turbulence and collected by the circular aperture.
 The solid line  shows the PDT for an atmospheric channel of  1 km length with the refractive index structure constant  $C_n^2{=}4{\times} 10^{-14}\,\text{m}^{-\frac{2}{3}}$ (Rytov parameter $\sigma_R^2{=}1.7$).
 The dashed line corresponds to a  channel of 2 km length with $C_n^2{=}3{\times} 10^{-15}\,\text{m}^{-\frac{2}{3}}$ ($\sigma_R^2{=}0.46$).
 The dot-dashed line  represents the PDT for a 3 km channel with $C_n^2{=}3{\times} 10^{-15}\,\text{m}^{-\frac{2}{3}}$ ($\sigma_R^2{=}0.96$).
 The additional  extinction losses of 1 dB/km are also included according to \cite{Elterman}.
}
\end{figure}

	 Let us firstly discuss  the case of weak  and weak-to-moderate  optical turbulence and calculate the corresponding  PDTs  for different propagation lengths and values of the structure constant $C_n^2$ by using the proposed approach.
	 The procedure of numerical evaluation of the PDTs is given in Appendix~\ref{app:PDTeval}.
	 We consider the optical beam, with an initial beam-spot size of $W_0{=}2$ cm at $\lambda{=}800$~nm, transmitted through the turbulence and collected by a  circular aperture with  radius $a{=}4$~cm.
	 The beam experiences beam wandering, characterized by the variance (\ref{SigmaBW}), and short-term beam broadening (\ref{Wst}).
	 The shape of the PDT is strongly influenced by the relative values $\sigma_{\mathrm{bw}}/a$ and   $W_{\mathrm{ST}}/a$  as well as by the Fresnel parameter $\Omega$ and the propagation length $L$ (see Fig.~\ref{fig:Bayes1}).

	 An interesting observation concerns the shape of the obtained PDTs.
	 For the case of 3-km  propagation the PDT resembles the log-normal distribution and thus shows a behavior typical for the case of saturated fluctuations (cf. the dash-dotted line in Fig.~\ref{fig:Bayes1}).
	 The beam wandering in this case plays a minor role and the beam broadening is the main source of transmission losses.
	 The channel characteristics based on the Rytov parameter suggest that the optical turbulence is rather weak.
	 On the other hand, the optical turbulence for the  1-km link is moderate and has a larger Rytov parameter due to the stronger local turbulent fluctuations of the refractive index.
	 The PDT for the 1-km link resembles the smoothed log-negative Weibull distribution, which shows that  beam wandering and  beam broadening contribute  to  fluctuating losses in the channel (cf. the solid line in Fig.~\ref{fig:Bayes1}).
	This behavior suggests that  the pronounced  beam-wandering effect is typical for short propagation lengths and stronger  refractive-index fluctuations given by the value of the  $C_n^2$ parameter.
	The growth of the turbulence strength leads to the increase of the size of the largest possible turbulent inhomogeneities and hence to the increase of the probability for the beam to be deflected as a whole, i.e., to the increase of beam wandering. 
	The resulting PDT then resembles the log-negative Weibull distribution \eqref{Pbw}.
	 On the other hand, the growth of the propagation length enhances the beam broadening due to the cumulative contribution of each propagation segment to the diffraction-induced broadening.
	As a consequence, for the sufficiently large broadening the beam-wandering effect diminishes and the resulting PDT resembles the truncated log-normal distribution \eqref{lnTrunc}.
	 Therefore, the shape of the PDT is influenced by the interplay of various factors and cannot be estimated by considering only the strength of optical turbulence given by $\sigma_R^2$.

\begin{figure}[ht]
 \includegraphics[width=0.45\textwidth]{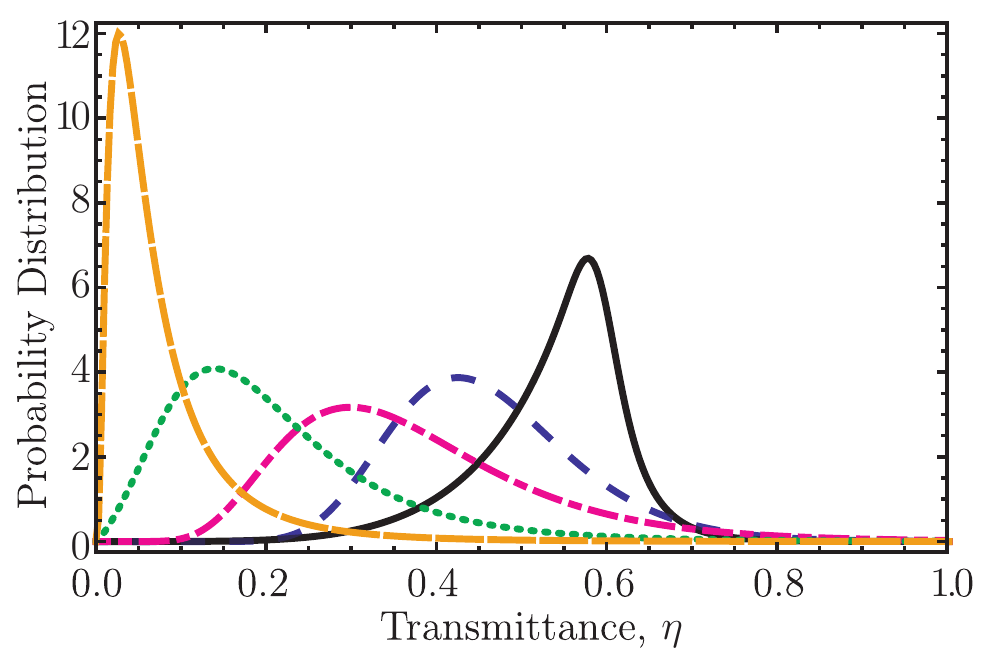}
 \caption{\label{fig:Bayes2}  The PDTs for a light beam propagating through the atmospheric turbulence with  the fixed refractive  index structure constant $C_n^2{=}4{\times} 10^{-14}\,\text{m}^{-\frac{2}{3}}$ and various propagation lengths:  1~km, solid line  ($\sigma_R^2{=}1.72$),  1.1~km, dashed line  ($\sigma_R^2{=}2.05$), 1.2~km, dash-dotted line  ($\sigma_R^2{=}2.41$); 1.5~km, dotted line  ($\sigma_R^2{=}3.60$); and  2~km,  long-dashed line  ($\sigma_R^2{=}6.14$).
 The other parameters are the same as in Fig.~\ref{fig:Bayes1}.
}\end{figure}

	 In Fig.~\ref{fig:Bayes2} the PDTs  are shown for the regimes of weak to moderate and moderate optical turbulence strength.
	 The optical beam propagates through  atmospheric quantum channels of different length and for a fixed refractive index structure constant $C_n^2{=4{\times}10^{-14}}\,\text{m}^{-\frac{2}{3}}$.
	 The beam and the receiver aperture  parameters are considered to be the same as in Fig.~\ref{fig:Bayes1}.
	 With the increase of the propagation distance and, consequently, with the increase of the Rytov parameter (\ref{RytovSigma}), the shape of the PDT changes from a distribution similar to the log-negative Weibull distribution to the log-normal form.
	 Note that  the transmission statistics changes quickly  with an increase of the Rytov parameter, due to increasing propagation distance.


\section{Beam tracking}\label{sec:Tracking}

	The goal of a tracking procedure is to increase the signal-to-noise ratio of the signal transmitted through the atmosphere  by mitigating the noise due to beam wandering~\cite{FriedYura}.
	The common  tracking system  involves  a position-sensitive sensor on the transmitter site, which detects the position variations of a reference beam (beacon) sent by the receiver.
	This sensor controls a fast-steering mirror that adjusts the source, aiming at its alignment with the receiver aperture center  (cf., e.g., Ref.~\cite{Ma2012, Yin2017, Kaushal}).
	The partial or complete mitigation of beam wandering results in the modification of the PDT that describes the atmospheric channel.
	In this section we derive the PDT for beam-tracking scenarios.

	The separation of the beam-wandering contribution from those induced by beam shape distortions  in Eq.~(\ref{Bayesian}) allows us to obtain the PDT in the case when the beam-tracking procedure is applied.
	The beam tracking  is aimed to minimize the beam deflection distance $r_0$ by continuous tracking of the instantaneous  position of the beam centroid and by the proper adjustment of the beam centroid relative to the aperture center.
	This adjustment is performed by shifting a fast steering mirror that targets the signal beam.
	Effectively this procedure  results in a decrease of the  variance $\sigma_{\mathrm{bw}}^2$ defined in Eq. (\ref{SigmaBW}).
	We also note that the following consideration is applicable also to situations when an additional beam jitter is present due to mechanical vibrations at the transmitter and/or receiver cites~\cite{Arnon}.
	In this case the variance  of distribution (\ref{BWprob}) includes contributions both from atmospheric beam wandering and from the vibrational  jitter.

	Let us define the beam-wandering variance after application of the tracking procedure as
	      \begin{align}\label{delta}
	      \Delta^2=\sigma_{\mathrm{bw}}^2-\sigma_{\mathrm{tr}}^2,
	      \end{align}
	where  $\sigma_{\mathrm{tr}}^2\in[0,\sigma^2_{\mathrm{bw}}]$ is the variance of the displacements of the beam centroid  due to the tracking procedure that characterizes the pointing error.
	The value  of $\Delta^2$ varies from $0$ for the perfect beam tracking to $\sigma_{\mathrm{bw}}^2$ for the no-tracking scenario.
	The intermediate values of $\Delta^2$ correspond to  partial corrections of fluctuating losses due to beam wandering.

	In the case of perfect beam tracking $ \Delta^2=0$, the distribution function (\ref{BWprob}) reduces to a Dirac $\delta$ function.
	Performing the integration in  Eq.~(\ref{Bayesian}) in this limiting case, one obtains
		\begin{align}\label{perfectTr}
		\mathcal{P}^{(\mathrm{perf})}(\eta)=P(\eta|0).
		\end{align}
	Therefore,  for a perfect-tracking scenario the PDT coincides with the conditional probability (\ref{Pcond}) with $r_0{=}0$.
	In view of our approximation (\ref{Pcond}), this means that the situation with perfect beam tracking is described by the truncated log-normal distribution (\ref{lnTrunc}).
	We also note that Eq.~(\ref{perfectTr}) is satisfied for  very long propagation lengths since in this case the effect of beam wandering is not pronounced~\cite{Khmelevtsov, Mironov1976}.

	For the imperfect beam tracking, the PDT is obtained from Eqs.~(\ref{Bayesian}) and (\ref{Pcond}) as
		\begin{align}\label{PDTtr}
		\mathcal{P}^{(\mathrm{tr})}(\eta)&{=}\frac{1}{2\Delta^2}\int\limits_0^\infty\D r_0 r_0
		\exp\left[-\frac{r_0^2}{2\Delta^2}\right]\nonumber\\
		&\times \mathcal{P}(\eta;\mu_{r_0},\sigma_{r_0}).
		\end{align}
	To evaluate  the integral we   use polar coordinates and perform the integration over the angular coordinate.
	Here $\mathcal{P}(\eta;\mu_{r_0},\sigma_{r_0})$ is the truncated log-normal distribution (\ref{lnTrunc}).

	For the quantitative  analysis of  beam tracking  we use the corresponding exceedance function $\overline{\mathcal{F}}^{(\mathrm{tr})}(\eta)$,
		\begin{align}\label{ExceedanceTr}
		\overline{F}^{(\mathrm{tr})}(\eta)&=\int\limits_\eta^\infty\D\eta^\prime  \mathcal{P}^{(\mathrm{tr})}(\eta^\prime),
		\end{align}
	where $\mathcal{P}^{(\mathrm{tr})}(\eta)$ is given by Eq.~(\ref{PDTtr}).
	Since   many  adaptive quantum protocols use the postselection of transmission events with large values of $\eta$ (cf.~Ref.~\cite{Vasylyev2012}), the exceedance characterizes the feasibility of such detection procedures based on beam tracking.
	The explicit expression for $ \overline{F}^{(\mathrm{tr})}(\eta)$ is given in Appendix~\ref{app:exceedance}.

\begin{figure}[ht]
 \includegraphics[width=0.45\textwidth]{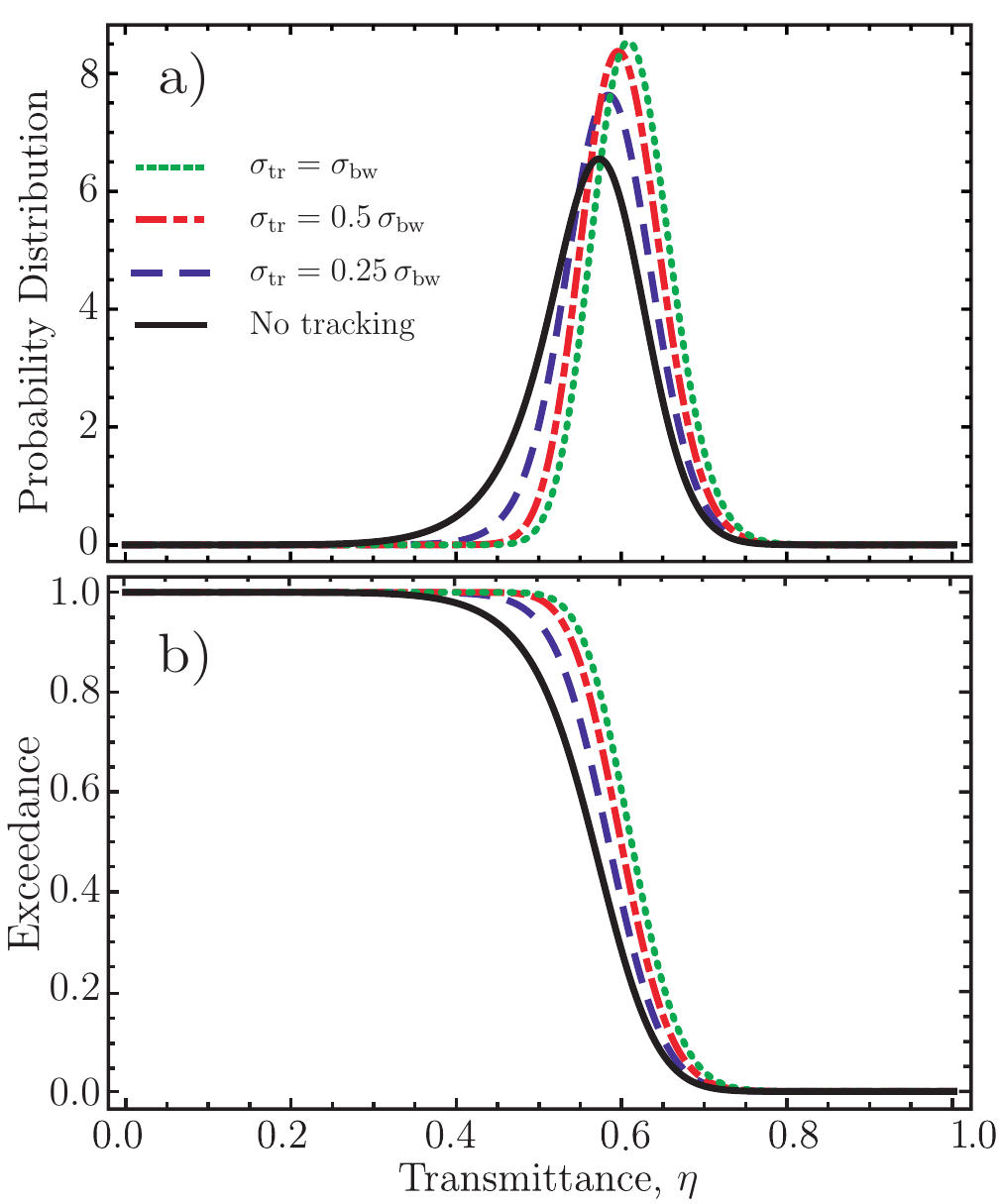}
 \caption{\label{fig:Tracking} Influence of the beam-tracking procedure on (a) the  PDT   and (b) the corresponding exceedance function, shown for  increasing values of the beam-tracking parameter: $\sigma_\mathrm{tr}{=}0$ (solid line), $\sigma_\mathrm{tr}{=}0.25\sigma_\mathrm{bw}$ (dashed line),  $\sigma_\mathrm{tr}{=}0.5\sigma_\mathrm{bw}$ (dash-dotted line), and $\sigma_\mathrm{tr}{=}\sigma_\mathrm{bw}$ (dotted line).
 The increase of  $\sigma_\mathrm{tr}$ results in a shift  of the maximum of the PDT or of the tail of the exceedance function to larger values of the transmittance $\eta$.
 This signifies the importance of beam tracking for protocols that utilize a  postselection of transmission events with a high transmittance.
 The case $\sigma_\mathrm{tr}{=}\sigma_\mathrm{bw}$ corresponds to perfect beam tracking.
 The  parameters of the beam and atmosphere correspond to those for the  solid curve in Fig.~\ref{fig:Bayes2}.
 }
\end{figure}

      Figure~\ref{fig:Tracking} shows the influence of the beam-tracking procedure on the statistical properties of the channel transmittance for a 1-km atmospheric link (Rytov parameter $\sigma_R^2{=}1.72$).
      The PDT of the considered atmospheric channel  and its  corresponding exceedance function are shown by solid lines.
      The application of the beam-tracking procedure  leads to a partial (dashed and dash-dotted lines) or full (dotted lines) mitigation of fluctuating losses caused by beam wandering.
      The tracking procedure  shifts the distribution tails  towards higher values of the transmittance.
      At the same time the exceedance function attains nonzero values for larger values of the transmittance $\eta$, with the growth of $\sigma_{\mathrm{tr}}$.
      This means that transmission scenarios with beam tracking better preserve nonclassical properties of the transmitted quantum light.
      As a consequence, beam-tracking  procedures improve the performance of quantum protocols with adaptive detection of the transmitted optical signal.

\begin{figure}[ht]
 \includegraphics[width=0.45\textwidth]{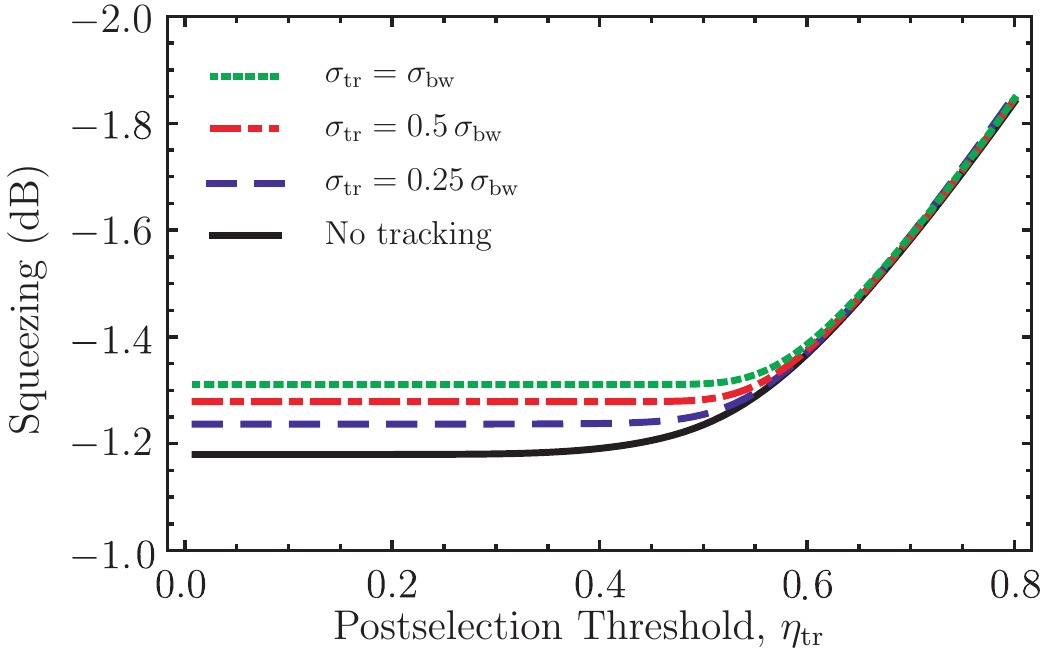}
 \caption{\label{fig:Squeezing}
Transmitted value of squeezing as a function of a postselection threshold $\eta_{\mathrm{min}}$.
The input light is  squeezed to $-2.4$ dB  and sent through a 1 km optical channel with $\sigma_R^2=1.72$.
It is detected by the receiver with the  application of both tracking and postselection procedures.
The solid line corresponds to the squeezing transmission without  beam tracking.
The corresponding PDTs and exceedance functions are shown in Fig.~\ref{fig:Tracking}.
}
\end{figure}

      We finally illustrate the application of a beam-tracking procedure in combination with postselection strategies for the preservation of nonclassical properties of transmitted quantum states of light.
      Figure~\ref{fig:Squeezing} shows the transmitted value of squeezing for an initial squeezing of $-2.4$~dB.
      The implemented postselection procedure selects the transmission events with transmittance values greater than the postselection threshold $\eta_{\mathrm{min}}$~\cite{Peuntinger, Vasylyev2016}.
      The application of the beam-tracking  procedure in general improves the detected squeezing, which is especially evident in Fig.~\ref{fig:Squeezing} for small values of the postselection threshold.
      This happens due to the larger signal-to noise ratio for the tracked beams in comparison to  non tracked signal detection.
      With the increase of $\eta_{\mathrm{min}}$ the value of the detected squeezing for tracked and non tracked beams is similar.
      Indeed, if $\eta_{\mathrm{min}}$ approaches the maximal possible transmittance values, the postselection procedure selects the transmission events when the beam centroid coincides or lies in the vicinity of the aperture center.
      As a consequence, the postselection with large values of $\eta_{\mathrm{min}}$ automatically restricts the detection to events with negligible beam wandering.
      In general, the applied beam-tracking procedure increases the feasibility or detection probability of squeezing in comparison to the non tracking case [cf.~Fig.~\ref{fig:Tracking} (b)].

\section{Decoy-state quantum communication through turbulent optical channels}\label{sec:decoy}

      In practical free-space quantum communication protocols the security analysis  of a transmitted key involves the description of the quantum communication channel.
      The signal losses caused by the propagation in the quantum channel are assigned to possible eavesdropper attacks and hence influence considerably  the security of the communication.
      In the Bennett-Brassard protocol  \cite{Bennett} both the channel losses and   contributions to the signal state from photon numbers higher than one lead to a security gap \cite{Huttner}.
      The decoy-state method~\cite{Hwang, Lo} was introduced in order to mitigate the security loophole connected with the multiphoton contributions of the source, which allowed one to enhance the protocol performance to a level comparable to that  with a perfect single-photon source.
      This scheme is based on the original Bennett-Brassardprotocol, where one communication party  sends the signal together with additional decoy states.
      The decoy states  are  used later for the detection of eavesdropping attacks.
      The rigorous analysis of the protocol security against the photon-number splitting attacks~ \cite{Ma, Skarani} or the Trojan-horse attacks~\cite{Tamaki} utilizes the knowledge of the transmittance properties of quantum communication channels.

      We consider the decoy-state protocol that utilizes an attenuated signal, a weak decoy-state (attenuated coherent quantum state), and the "empty" decoy state (vacuum quantum state) with the mean photon numbers $\mu_s$, $\mu_d$, and $\mu_v$, correspondingly.
      For the considered protocol the following conditions hold true ~\cite{Ma}:  $\mu_s{<}\mu_d{<}1$ and $\mu_v{=}0$.
      In Ref.~\cite{Manderbach} the successful implementation of this protocol was shown for the 144-km atmospheric channel between two Canary Islands.
      The theoretical analysis of the performance of the decoy-state protocol  that utilizes atmospheric quantum channels is given in \cite{Scott, Gruneisen, Lo2018} and the experimental realization is demonstrated in \cite{Liao2018}.
      The transmitter Alice encodes the pulses in a signal and two decoy states and sends them to the receiver Bob.
      The vacuum decoy state serves for the estimation of the  background noise yield.
      On the other hand, the combination of measurements of weak decoy states and vacuum decoystates  allows one to estimate the relevant parameters  for the single-photon components, including the yield and quantum-bit error rate (QBER).

      After random encoding of bits in the X or Z basis by Alice and Bob's measurements of transmitted bits in a randomly chosen X or Z  basis, the parties perform the sifting of the raw key, its error correction and privacy amplification.
      As a result, Alice and Bob  extract a shorter but more secure key.
      The lower bound for the averaged secure key rate is given by  \cite{Ma2006}
	      \begin{align}\label{KeyRate}
	      \mathcal{R}=\frac{1}{2}\int_0^1\D\eta & \mathcal{P}(\eta)\left\{Q_{1}^s(\eta)\left[1-h(e_1^{\mathrm{ph}})\right]\right.\nonumber\\
	      & \left.\qquad-Q_{\mu_s}(\eta) f\, h[E_{\mu_s}(\eta)]\right\},
	      \end{align}
      where the   averaging is performed over the  atmospheric transmittance with the corresponding PDT $\mathcal{P}(\eta)$.
      Here $h(x){=}-x\log_2x{-}(1{-}x)\log_2(1{-}x)$ denotes the binary entropy function and $f$ is the inefficiency of error correction.
      The gain  $Q_{\mu_s}=M_{\mathrm{click}}^s/N^s$ represents the ratio between the number $M_{\mathrm{click}}^s$ of events where Bob observes a click in his measurement device under the condition that Alice sent $N^s$ signals.
      The QBER $E_{\mu_s}{=}M_{\mathrm{error}}^s/M_{\mathrm{click}}^s$ is the ratio between the number of errors observed by Bob to the number of detected events.
      The one-photon gain for the signal state $Q_{1}^s$ can be estimated from the transmission characteristics of the signal and the weak decoy state as described in Ref.~\cite{Ma2006}, with the  lower bound given by
	      \begin{align}
		Q_1^s(\eta)&{=}\frac{\mu_s^2\,e^{-\mu_s}}{\mu_s\mu_d{-}\mu_d^2} \Bigl(Q_{\mu_d}(\eta)e^{\mu_d}\nonumber\\
		&-Q_{\mu_s}(\eta)e^{\mu_s}\frac{\mu_d^2}{\mu_s^2}{-}\frac{\mu_s^2-\mu_d^2}{\mu_s^2}Y_0\Bigr).
	      \end{align}
      Finally, $e_1^{\mathrm{ph}}$ in Eq.~(\ref{KeyRate}) denotes the phase error rate, whose upper bound for finite key length can be estimated as described in Ref.~\cite{Fung}.

\begin{figure}[ht]
 \includegraphics[width=0.45\textwidth]{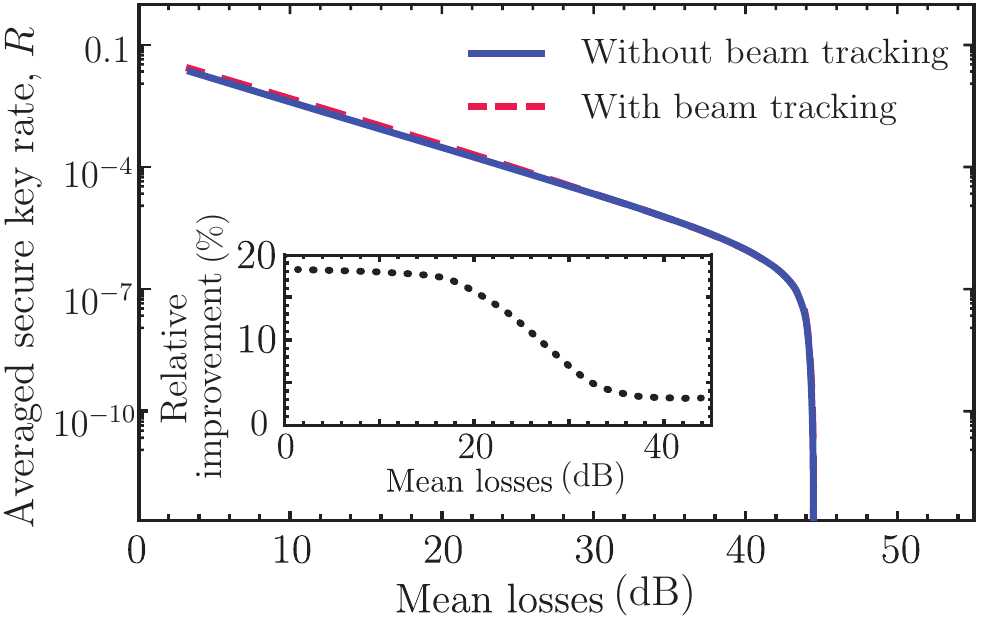}
 \caption{\label{fig:Bayes3} Averaged secure key rate for the two-decoy-state quantum key distribution in an atmospheric  quantum channel as a function of mean channel losses without (solid line) and with  beam tracking (dashed line).
 The signal  with the mean photon number $\mu_s{=}0.27$ at $\lambda{=}800$~nm is mixed with a weak decoy state  ($\mu_d{=}0.39$) and an empty decoy state  ($\mu_v{=}0$).
 It is sent through atmospheric links of various lengths but equal refractive-index structure function.
 Other beam and aperture   parameters are the same as in Fig.~\ref{fig:Bayes1}.
 The additional 1-dB/km losses due to  absorption are also taken into account.
 The inset shows the relative improvement given in percents [cf. Eq.~\eqref{RelativeImpr}] of the average secure rate by applying the beam-tracking procedure.
}
\end{figure}

      For the signal or decoy states transmitted through the atmospheric quantum channel the gain and QBER values can be written as
	      \begin{align}\label{yield}
	      Q_{\mu_i}(\eta)=Y_0+1-e^{-\eta_d\,\eta\, \mu_i}
	      \end{align}
      and
	      \begin{align}\label{QBER}
	      E_{\mu_i}(\eta)=\frac{1}{ Q_{\mu_i}(\eta)}\left\{\frac{1}{2} Y_0+e_{\mathrm{det}}\left[Q_{\mu_i}(\eta){-}Y_0\right]\right\},\quad i{=}s,d,
	      \end{align}
      respectively.
      Here $\eta$  is the fluctuating transmittance of the turbulent atmosphere and $\eta_d$ is the transmittance affected by deterministic losses, such as losses in key generation and detection modules, atmospheric absorption, etc.
      We use the following values for parameters in Eqs.~(\ref{yield}) and (\ref{QBER}): the zero-photon yield  $Y_0=1.7\times 10^{-6}$, the misalignment error rate, $e_{\mathrm{det}}{=}0.01$, the inefficiency of error correction, $f=1.2$, and the mean photon numbers of the signal and the  weak decoy field  are $\mu_s{=}0.27$ and $\mu_d{=}0.39$, respectively.

      In order to illustrate the applicability of the present approach for the simulation of atmospheric communication links, we calculate the averaged secure key rate (\ref{KeyRate}) as a function of mean channel losses.
      We consider  atmospheric channels of different lengths but characterized by the same  refractive index structure constant, $C_n^2{=}4{\times} 10^{-14}\,\text{m}^{-\frac{2}{3}}$.
      The statistics of the particular channel are derived by repeating the transmission simulation, governed by the corresponding PDT in Eq.~(\ref{Bayesian}), 10000 times for every propagation length.
      For every simulated value of the transmittance $\eta$, the gain (\ref{yield}) and QBER (\ref{QBER}) functions are calculated and substituted into Eq.~(\ref{KeyRate}).
      Finally, the  average value of the secure key rate is then calculated.
      Figure~\ref{fig:Bayes3} shows the results of the simulations for the two-decoy state protocols operating at  $\lambda{=}800$~nm.
      The length of the raw key is assumed to be sufficiently large, such that  the phase error rate is approximately equal to the QBER, i.e.,  $e_1^{\mathrm{ph}}{\approx}E_{\mu_{s}}$ (see Ref.~\cite{Fung}).

      Figure~\ref{fig:Bayes3}  shows that the averaged secure rate decreases with the increase of the propagation length and hence of the mean losses.
      Around 45 dB of mean losses, the secure key rate degrades significantly and the secure communication using the two-decoy state protocols becomes impossible.
      This corresponds to 8.25-km propagation distance for the considered atmospheric and beam parameters.

      Figure~\ref{fig:Bayes3}  compares the average secure key rate for the cases with (dashed line) and without (solid line) beam tracking.
      To clarify  the comparison  the relative improvement
	    \begin{align}\label{RelativeImpr}
	     \mathcal{I}=1-\frac{\mathcal{R}(\Delta{=}\sigma_{\mathrm{bw}})}{\mathcal{R}(\Delta=0)},
	    \end{align}
      i.e., the characteristics of how the beam-tracking procedure improves the security of the decoy-state protocol, is shown in the inset of Fig.~\ref{fig:Bayes3}.
      Here $\mathcal{R}(\Delta)$ is the average key rate \eqref{KeyRate} for the protocol with beam tracking which is calculated by using the corresponding PDT given by Eq.~\eqref{PDTtr}.
      Under the particular atmospheric and propagation conditions, the beam-tracking procedure improves  the security for short propagation distances (low mean losses).
      For large mean losses, the transmission losses due to beam broadening dominate the beam-wandering effect and hence the tracking procedure has  minor influence in the region of large mean losses.
      This dependence can be explained by the observation that for large-$L$ values the beam-wandering variance saturates~\cite{Mironov1976} and hence the compensation of beam wandering does not significantly improve the channel transmission.


\section{Summary and Conclusions} \label{sec:Summary}

	We introduced a technique for deriving the probability distribution for the atmospheric transmittance based on the idea to separate the contributions of beam wandering from those of beam-spot distortions.
	This technique is based on the law of total probability.
	We show that the probability distribution for the channel transmittance depends on classical optical field-correlation functions of second and fourth order.
	The advantage of the present method is that the first two moments of the probability distributions, i.e., the mean channel transmittance and the mean-square transmittance, coincide automatically with those which are calculated from  first principles.
	This important property allows one to describe  atmospheric quantum channels also in situations when the elliptic-beam approximation~\cite{Vallone2016} does not apply and other proper methods do not exist, e.g. for  very long propagation distances.
	However, the main limitation of the present approach is the computational complexity of fourth-order field-correlation functions, for general turbulence conditions.
	To overcome this problem, we have used the approximation of weak beam wandering.
	The latter does not require knowledge of the whole fourth-order field-correlation function, but only two related integral quantities, the mean-squared transmittance and the beam-wandering variance.

	The separation of beam-wandering from beam-distortion contributions allows one to incorporate the description of atmospheric quantum channels for propagation scenarios with the often applied beam-tracking technique.
	Beam tracking mitigates losses due to beam-wandering and hence it helps to preserve quantum properties of the transmitted light.
	Further improvement of the quantum-channel performance could be achieved by combining beam tracking with the postselection procedures, where the transmission events characterized by larger values of the transmittance are preferably used.
	We have derived the corresponding exceedance function that serves for the characterization of such postselection strategies with beam tracking.
	
	Finally, we have analyzed the security of decoy-state quantum key distribution protocols in atmospheric channels with strong turbulence.
	We have found that the mitigation of turbulence-induced beam wandering improves the  secure key rate for channels with small mean losses or short propagation lengths.
	However, in the region of large mean losses or long propagation distances the beam tracking has minor influence on the secure key rate.
	The degradation of the security of the protocol occurs for the same level of mean losses for protocols with and without the beam-tracking procedure.
	Hence, we conclude that the beam tracking does not always improve quantum communication.

\acknowledgments  The authors gratefully acknowledge useful discussions with O.~Chumak and V.~Usenko.
The work was supported by the Deutsche Forschungsgemeinschaft through Project No. VO 501/21-2 and the European Union's Horizon 2020 Research and Innovation program under Grant Agreement No. 665148 (QCUMbER).

\appendix


\section{Parameters}\label{app:parameters}

	In this appendix we give explicit formulas for the functions  $\eta_0$, $R$, and $\lambda$ that  occur in Eq.~(\ref{Pbw}).
	For further details we refer to Ref.~\cite{Vasylyev2012}.
	These functions appear as parameters in the approximated dependence of the aperture  transmittance of Gaussian beams  on the deflection length $r_0$ between the beam centroid and the aperture center [cf. Eq.~(\ref{CondEta1})].
	For the circular aperture with radius $a$, the maximal transmittance of a Gaussian beam with  the beam-spot width $W_{\mathrm{ST}}$ is obtained when the beam centroid position coincides with the aperture center and reads
		\begin{align}
		\eta_0=1-\exp\left[-2\frac{a^2}{W_{\mathrm{ST}}^2}\right].
		\end{align}
	The scale and shape parameters of the approximation (\ref{CondEta1}) are
		\begin{align}
		& R=a\Bigl[\ln\Bigl(2\frac{\eta_0}{1-\exp\bigl[-4\frac{a^2}{W_{\mathrm{ST}}^2}\bigr]}\I_0\bigl(4\frac{a^2}{W_{\mathrm{ST}}^2}\bigr)\Bigr)\Bigr]^{-\frac{1}{\lambda}},
		\end{align}
		\begin{align}
		\lambda&=8\xi^2\frac{e^{-4\frac{a^2}{W_{\mathrm{ST}}^2}}\I_1\bigl(4\frac{a^2}{W_{\mathrm{ST}}^2}\bigr)}{1-\exp\bigl[-4\frac{a^2}{W_{\mathrm{ST}}^2}\bigr]\I_0\bigl(4\frac{a^2}{W_{\mathrm{ST}}^2}\bigr)}\nonumber\\
		&\times \left[\ln\left(2\frac{\eta_0 }{1-\exp\bigl[-4\frac{a^2}{W_{\mathrm{ST}}^2}\bigr]\I_0\bigl(4\frac{a^2}{W_{\mathrm{ST}}^2}\bigr)}\right)\right]^{-1}.
		\end{align}
	Here  $\I_n(x)$ is the modified Bessel function of $n$th order.


\section{Evaluation of the PDT}\label{app:PDTeval}

	In this appendix we outline the procedure of the numerical evaluation of the PDT (\ref{Bayesian}).
	This procedure involves the following steps.
		\begin{itemize}
		  \item[(i)]  Calculate numerically the parameters $\langle\eta\rangle$, $\langle\eta^2\rangle$, $\sigma^2_{\mathrm{bw}}$, and $W_{\mathrm{ST}}^2$ given by Eqs.~(\ref{MEta}), (\ref{MEta2}), (\ref{SigmaBW}), and (\ref{Wst}), respectively.
		  The required correlation functions are given by Eqs.~(\ref{Gamma2Calc}) and (\ref{Gamma4Calc}) for the focused Gaussian beam.
		  \item[(ii)] The numerical integration in Eq.~(\ref{Bayesian}) can be performed with a Monte Carlo method.
		  For this purpose one should simulate  the $N$ values  of the vector $\boldsymbol{r}_0$.
		  The obtained values $[r_0]_i$, $i=1,...,N$, are normally distributed according to Eq.~(\ref{BWprob}) with zero mean and variance (\ref{SigmaBW}).
		  The simulated values of $\boldsymbol{r}_0$ are substituted into Eq.~(\ref{muApprox}) and the set of $N$ values $[\mu_{r_0}]_i$  is formed.
		  Finally, the parameter $\sigma_{r_0}$ is calculated according to Eq.~(\ref{sigmaApprox}).
		  \item[(iii)] The PDT can be estimated from the simulated values of $[\mu_{r_0}]_i$ as
		      \begin{align}
		      \mathcal{P}(\eta)\approx\frac{1}{N}\sum\limits_{i=1}^N \mathcal{P}(\eta;[\mu_{r_0}]_i,\sigma_{r_0}),
		      \end{align}
		  where $\mathcal{P}(\eta;\mu,\sigma)$ is the truncated log-normal distribution given by Eq.~(\ref{lnTrunc}).
		  \item[(iv)] The mean value of any physical quantity that is a function of the transmittance can be estimated as
		      \begin{align}
		      \langle f(\eta)\rangle\approx\frac{1}{N}\sum\limits_{i=1}^N \langle f[\eta(\eta)]\rangle_i,
		      \end{align}
		  where $\langle f(\eta)\rangle_i$ is obtained from the log-normal distribution with parameters $[\mu_{r_0}]_i$, $\sigma_{r_0}$.
		\end{itemize}
	If the partial beam tracking procedure is applied, the evaluation of the PDT is performed with the replacement $\sigma^2_{\mathrm{bw}}\rightarrow\Delta^2$ [cf.~Eq.~(\ref{delta})].


\section{Beam-tracking exceedance}\label{app:exceedance}

	In this Appendix we give the explicit expression for the exceedance functions under the condition that beam tracking is performed.
	Inserting Eq.~(\ref{PDTtr}) into Eq.~(\ref{ExceedanceTr}) and performing the integration over $\eta$ yields
		\begin{align}\label{Exceedance1}
		&\overline{\mathcal{F}}^{(\mathrm{tr})}(\eta)=\frac{1}{2\Delta^2}\int\limits_0^\infty\D r_0 r_0 \exp\left[-\frac{r_0^2}{2\Delta^2}\right]\nonumber\\
		&\times\frac{1}{2\mathcal{F}^{(\mathrm{tr})}(1)}\left[\mathrm{erf}\left(\frac{\mu_{r_0}}{\sqrt{2}\sigma_{r_0}}\right)-\mathrm{erf}\left(\frac{\ln\eta+\mu_{r_0}}{\sqrt{2}\sigma_{r_0}}\right)\right],
		\end{align}
	where
		\begin{align}
		\mathcal{F}^{(\mathrm{tr})}(1)&=\int_0^1\D\eta \mathcal{P}(\eta;\mu_{r_0},\sigma_{r_0})\nonumber\\
		&=\frac{1}{2}\left[1+\mathrm{erf}\left(\frac{\mu_{r_0}}{\sqrt{2}\sigma_{r_0}}\right)\right].
		\end{align}
	Equation (\ref{Exceedance1}) can be alternatively written  as
		\begin{align}\label{Exceedance2}
		\overline{\mathcal{F}}^{(\mathrm{tr})}(\eta)&=\frac{1}{2\Delta^2}\int\limits_0^\infty\D r_0 r_0 \exp\left[-\frac{r_0^2}{2\Delta^2}\right]\nonumber\\
		&\qquad\times\left[1- \frac{1+\mathrm{erf}\left(\frac{\ln\eta+\mu_{r_0}}{\sqrt{2}\sigma_{r_0}}\right)}{1+\mathrm{erf}\left( \mu_{r_0}/\sqrt{2}\sigma_{r_0}\right)}\right].
		\end{align}	
	The remaining integration in Eqs.~(\ref{Exceedance1}) and (\ref{Exceedance2}) should be performed numerically.
	The corresponding beam-tracking correlation functions are given by
		\begin{align} \label{TrackedGamma2}
		\Gamma_2^{(\mathrm{tr})}(\boldsymbol{r},L){=}\exp\left[-\frac{\sigma_{\mathrm{tr}}^2}{2}\Delta_{\boldsymbol{r}}\right]\Gamma_{2}(\boldsymbol{r},L),
		\end{align}
	and
		\begin{align}\label{TrackedGamma4}
		\Gamma_4^{(\mathrm{tr})}(\boldsymbol{R},\boldsymbol{\rho},L){=}\exp\left[-\sigma_{\mathrm{tr}}^2
		\Delta_{\boldsymbol{R}}\right]\,\Gamma_{4}(\boldsymbol{R},\boldsymbol{\rho},L).
		\end{align}


\end{document}